\documentclass[12pt]{article}
\usepackage{epsfig, axodraw}
\usepackage [small] {caption}
\newcommand{\bq}{\begin{eqnarray}}
\newcommand{\eq}{\end{eqnarray}}
\newcommand{\ra}{\rightarrow}
\newcommand{\ov}{\overline}

\newcommand{\ak}{\alpha_s(k^2)}
\newcommand{\az}{\frac{\ak}{\alpha_s(\ov M_Q^2)}}

\begin{document}
\begin{center}
\bf{Is the BELLE result for the cross section} $\sigma (e^+e^-\ra J/\psi+
\eta_c)$ {\bf a real difficulty for QCD\,?}
\end{center}
\vspace{1cm}
\begin{center}{\bf A.E. Bondar, V.L. Chernyak} \end{center}
\begin{center}Budker Institute of Nuclear Physics, Novosibirsk, Russia
\end{center}

\vspace{3cm}
\begin{center}Abstract \end{center}

It is shown that difficulties in reconciling the values of the cross section
$\sigma(e^+e^-\ra J/\psi+\eta_c)$ measured at BELLE and calculated within 
non-relativistic QCD (NRQCD) are caused not by some misinterpretation of the 
data or other exotic explanations, but by poor applicability of NRQCD to 
such processes. 

We use general theory of hard exclusive processes in QCD together with  more 
realistic models of charmonium wave functions, and show that the BELLE result
can be naturally explained.

\newpage
{\bf 1.} A surprisingly large rate for hard exclusive processes of 
the type $e^+ e^-\to J/\psi+\eta_c$ observed at BELLE~\cite{BE1} still
remains unexplained. In this experiment the process $e^+e^-\to J/\psi+X$ was
studied. The cross-section of $e^+ e^-\to J/\psi+\eta_c$ was then extracted 
from the number of events in the $\eta_c$ peak in the mass 
spectrum of the system recoiling against the reconstructed $J/\psi$.

In a recent upgrade of the BELLE analysis with a data sample of 
$155\,$fb$^{-1}$ \cite{BE2},\cite{P} the cross-section of the  $e^+ e^
-\to J/\psi+\eta_c$ process has been found equal to $(25.6\pm2.8\pm3.4)\,
{\rm fb}/Br(\eta_c > 2\, {\rm charged})$. BELLE also performed simultaneous 
fits to the production and helicity angle distributions. The measured angular 
and helicity distributions for $J/\psi+\eta_c$ have the general form 
$(1+\alpha\cos^2{\theta})$ and are consistent with the expectations for 
production of this final state via a single virtual photon, $\alpha_{prod}
= \alpha_{hel} = +1$.

From the theoretical side, this cross section was calculated in \cite{BL}
within the NRQCD approach, and much smaller value $\simeq 2.3\,{\rm fb}$ was
obtained. This large discrepancy initiated further studies, both 
experimental and theoretical. Various explanations were proposed, e.g. that
the two-photon production of $(J/\psi+J/\psi)$ can be significant and 
can imitate those of $J/\psi+\eta_c\,\,$ \cite{BBL},
\footnote{\hspace{2mm} 
Later, this possibility was excluded by BELLE~\cite{BE2}.
}
or even more exotic variants of the scalar gluonium production 
(for its mass happily coinciding with the charmonium energy levels) 
\cite{BGL}, etc. 

\vspace{0.5cm}

The main purpose of this paper is to show that the origin of the discrepancy
is due to a poor accuracy of NRQCD when applied to such processes. The
main reason is that the charm quark is not sufficiently heavy and, as a 
result, the charmonium wave functions are not sufficiently narrow for a
reasonable application of NRQCD to the description of charmonium production.
And, as usual, hard exclusive processes are particularly sensitive to the
widths of hadron wave functions. Below we describe the properties of the 
model wave functions of charmonium (which we consider as more realistic in 
comparison with the extreme $\delta$-function-like wave functions of NRQCD),
and show that the value of $\sigma(e^+e^-\ra J/\psi+\eta_c)$ measured at 
BELLE can be naturally obtained.

{\bf 2.} The cross section of the process $e^+e^-\ra \gamma^*\ra J/\psi_{
\perp}(p_1)+\eta_{c}(p_2)$ has the standard form:
\bq
\sigma \biggl (e^+e^-\ra \gamma^*\ra J/\psi_{\perp}(p_1)+\eta_c(p_2)\biggr )
=\frac{\pi
\alpha^2}{6}\Biggl (\frac{|\vec p|}{E}\Biggr )^3 Q^2_c\,|F_{VP}(s)|^2\,,
\eq
where $(|\vec p|/E)^3$ is the P-wave phase space factor and $Q_c=2/3$ 
is the charm quark charge, while the form factor is defined as:
\bq
\langle \psi_{\perp}(p_1),\eta_c(p_2)|J_{\mu}(0)|0\rangle=
\epsilon_{\mu\nu\rho\sigma}e_{\perp}^{\nu}p_1^{\rho}p_2^
{\sigma}\,F_{VP}(s)\,.
\eq
Since there is one form factor only, the angular distribution is pure 
kinematical: $\sim (1+\cos^2 \theta)$.

We will need only the asymptotic form of $F_{VP}(s)$ because  
$s=M^2_{\Upsilon (4S)}\simeq 112\,{\rm GeV^2}$ in the BELLE experiment.
\vspace{1.0 cm}

General theory of hard exclusive processes in QCD has been developed in 
\cite{cz1},\cite{cz2} (see \cite{CZ} for a review). In \cite{cz1} it was 
obtained that at large $s=(p_1+p_2)^2$ the leading power term of the general 
two-hadron form factor has the following behaviour (up to logarithmic 
corrections):
\bq
\langle H_1(p_1,s_1,\lambda_1;\,
H_2(p_2,s_2,\lambda_2)|J_{\lambda}|0\rangle\sim
\biggl (\frac{1}{\sqrt s}\biggr )^{|\lambda_1+\lambda_2|+2n_{min}-3}\,.
\eq
Here: $H_1$ and $H_2$ are any two hadrons with momenta $p_{1,2}$\,, spins
$s_{1,2}$ and helicities $\lambda_{1,2}$ in c.m.s.,\, the current helicity 
(here, the projection of the photon spin onto direction ${\vec p}_1$) is:
$\lambda=(\lambda_1-\lambda_2)=0,\,\pm 1$. $n_{min}$ is the minimal number of
point-like constituents (quarks or gluons) in these hadrons, $n_{min}=2$ for 
mesons and $n_{min}=3$ for baryons. It is seen that the asymptotic behaviour 
is independent of hadron spins, but depends essentially on their helicities.

For the process considered, $e^+e^-\ra \gamma^*\ra J/\psi_
{\perp}(p_1)+\eta_{c}(p_2)$, the $J/\psi$-meson is transversely 
polarized, i.e. has helicities $|\lambda_1|=1$ only. 
So, the matrix element in eq.(3) behaves as $\sim 1/s$. Since in eq.
(2) $e_{\perp}\sim 1$, while $p_1\sim p_2\sim \sqrt s$, all that results in\,
$\,F_{VP}(s)\sim 1/s^2$.

The leading term of $F_{VP}(s)$ is given by four similar 
diagrams, one of which is shown in Fig.1. Its explicit
form will be given below in eq.(9,10), but we describe first the properties
of the meson wave functions entering eq.(10).

\begin{figure}
\centering
\includegraphics[width=0.3\textwidth] {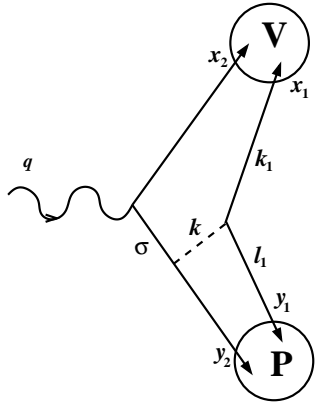}
\caption{One of the four similar diagrams for the form factor
$F_{J/\psi\,\eta_c}$.
}
\label{Fig.1}
\end{figure}

\vspace{0.5cm}

{\bf 3.} The twist 2 and twist 3 light cone wave functions $V_{i}(x)$ and 
$P_{i}(y)$ of the $^3 S_1$ and $^3 S_0$ states of quarkonium
made of the heavy ${\ov Q}Q$ quarks are defined in the standard way 
(see e.g. \cite{CZ}), we neglect higher twist wave functions 
giving only power corrections in comparison with those written below. For
simplicity, we also 
do not distinguish $M_{V}$ from $M_{P}$ and use $\ov M$
which can be taken as an appropriate average, for instance: ${\ov M}=(3M_
{V}+M_{P})/4$. 

For the $V$-meson with the helicity $\lambda$:
\footnote{\hspace{2mm}
We use the conventions:
\bq
\sigma_{\mu\nu}=[\gamma_{\mu},\gamma_{\nu}]/2,\quad \gamma_5=(\gamma_
5)_{Bj-Dr},\quad \widehat p=p_{\mu}\gamma^{\mu}=p_o\gamma_o-
{\vec p}\,{\vec \gamma}. \nonumber
\eq
}
\bq
{\langle V_{\lambda}(p)|{\ov Q}_{\beta}(z)\,Q_{\alpha}(-z)|0\rangle}_{\mu^*}=
\frac{f_{V}\ov M}{4}\int^1_o dx_1\,e^{i(pz)(x_1-x_2)}
\biggl \{ {\widehat e}_{\lambda}\,V_{\perp}(x)+\nonumber
\eq
\bq
+{\widehat p}\,\frac{(e_{\lambda}z)}{(pz)}\,{\widetilde V}(x)+
f^{t}_{v}(\mu^*) (\sigma_{\mu\nu} e_{\lambda}^{\mu}\, p^{\nu})\,V_{T}(x)+
f^{a}_{v}(\mu^*)(\epsilon_{\mu\nu\alpha\beta}\gamma_{\mu}\gamma_5\,e_{\lambda}
^{\nu}\,p^{\alpha}z^{\beta})\, V_{A}(x) \biggl \}_{\alpha\beta}.
\eq

Taking $\lambda=0$ in eq.(4) and using
${\ov M}e^{\mu}_{\lambda=0}\ra p_{\mu}$ for large $p$, one obtains the 
standard definition of the leading twist wave function $V_L(x)$ of 
longitudinally polarized vector meson \cite{CZ}:
\bq
\langle V_{\lambda=0}(p)|{\ov C}(z)\gamma_{\mu}C(-z)|0\rangle =f_{V}
\int^1_o dx_1\,e^{i(pz)(x_1-x_2)}V_{L}(x), 
\eq
so that: ${\widetilde V}(x)=V_{L}(x)-V_{\perp}(x)$.

$x_1$ and $x_2=(1-x_1)$ in eqs.(4,5) are fractions of the meson momentum 
$p^+=(E+p_z)\equiv q_o$ carried by quarks at large $p_z$, i.e.
$[+\,,\perp,\,-]$ components of the quark momentum are: $k_1=
[x_1 q_o,\,0_{\perp},\,M_Q^2/x_1 q_o]$ (and similarly for three other quarks), 
and we neglect the quark transverse momentum inside the heavy quarkonium in 
comparison with its mass. 

For the $P$-meson:
\bq
{\langle P(p)|{\ov Q}_{\beta}(z)\,Q_{\alpha}(-z)|0\rangle}_{\mu^*}=
i\frac{f_{P} {\ov M}}{4}\int^1_0 dx_1 e^{i(pz)(y_1-y_2)}\biggl \{ 
\frac{\widehat p\,\gamma_5}{\ov M}\,P_{A}(y)-\nonumber
\eq
\bq
-f_{p}^{p}(\mu^*)\,\gamma_5\,P_{P}(y)+f_{p}^t(\mu^*)\, 
(\sigma_{\mu\nu}\,p^{\mu}\, z^{\nu})\,P_{T}(y)\biggr \}_{\alpha\beta} \,.
\eq

The wave functions and operators entering eqs.(4-6) are defined at the soft 
scale $\mu^*,\,\,\,\mu^*\ll q_B$, where 
$q_B=C_F\alpha_s(\mu\simeq q_B)M_Q^*/2$ is the Bohr momentum of the heavy 
quark in the quarkonium. 

All wave functions in eq.(4-6) are symmetric under: $x_1\leftrightarrow x_2,
\, y_1\leftrightarrow y_2$ and normalized: $\int_0^1 dx_1\,V_i(x)=1,\,\, i=
\perp,\,L,\,T,\,A;\quad \int_o^1 dy_1\,P_i(y)=1,\,\,i=A,\,P,\,T\,.$ 

The values of dimensionless constants in the above formulae 
follow directly from the exact QCD equations of motion: $i{\hat D}Q=
M_Q^{*} Q$ (see e.g. ch.9 and Appendix C in \cite{CZ}; here $M_Q^{*}$ 
is the "soft" mass of the heavy quark, e.g. the appropriately
defined perturbative pole mass)
\footnote{\hspace{2mm}
$\Delta_s$ in eq.(7) originates from the matrix element: $\langle V_{\lambda}
(p)|{\ov Q}(\stackrel{\rightarrow}{iD_\mu}-\stackrel{\leftarrow}{iD_\mu})Q|0
\rangle_{\mu^*}\equiv \Delta_s f_V{\ov M}^2 e_{\mu}^{\lambda}$. It is $O(v^2)
$ parametrically, and up to $O(v^4)$: $\Delta_s\simeq (4M_Q^{*2}-{\ov M}^2)
/6{\ov M}M_Q^*$. For the c-quark pole mass $M_c^*=1.65\,{\rm GeV}: \,\, 
\Delta_s\simeq 0.05$, and will be neglected in what follows.
}
:
\bq
f_{p}^p(\mu^*)=\frac{\ov M}{2M_Q^*}\,,\quad f^t_{v}(\mu^*)=\biggl (\frac{2M_
Q^*}{\ov M}-\Delta_s \biggr ),
\eq
\bq
f_{v}^a(\mu^*)=\frac{1}{2}\biggl ( 1-f_{v}^t(\mu^*)\frac{2M_Q^*}{\ov M}\biggr
),\quad f^t_{p}(\mu^*)=\frac{1}{3}\biggl (1- \frac{(2M_Q^*)^2}{\ov M^2}
\biggr ).\nonumber
\eq

However, the appropriate scale\, $\mu\,$ for the wave functions entering the 
hard form factor $F_{VP}(s)$ will be $\mu^2\simeq k^2$  rather than $\mu^{*2}
$, see Fig.1. For simplicity, we neglect in what follows the complicated but 
slow logarithmic evolution of (normalized) wave function forms, and will 
account only for the overall renormalization factors of the local tensor and 
pseudoscalar currents and for running of the quark mass. Unlike the larger 
soft mass $M_Q^*$, the mass ${\ov M}_Q$ entering below the perturbative 
logarithmic renormalization factors and couplings $f_{v,p}^i(k^2)$ 
will be the "hard" 
mass, e.g. ${\ov M}_Q=M_Q^{\ov{MS}}(\mu=M_Q^{\ov{MS}})$, which is smaller 
because a part of the heavy quark self-energy in the interval $(1/{\ov M
}_Q)< R < (1/\mu^*)$ is now excluded from $M^*_Q$. So, we use in eq.(10)
(the renormalization factors $Z_p,\,Z_t$ and $Z_m^k$ are given below):
\footnote{ \hspace{2mm}
The function $P_T(x)$ gives no contribution to the form factor $F_{VP}(s)$.
} 
\bq
f_{p}^p(k^2)= \frac{\ov M}{2{\ov M}_Q}Z_{p},\quad f_{v}^t(k^2)=
\frac{2{\ov M}_Q}{\ov M}Z_t,\quad f_{v}^a(k^2)= \frac{1}{2}\biggl 
(1-Z_t Z_m^k\frac{4{\ov M}_Q^2}{{\ov M}^2 } \biggr )\,.
\eq

\vspace{0.5cm}

{\bf 4.} The leading contribution to the form factor $F_{VP}(s)$ is 
calculated in a standard way (see ch.9 in \cite{CZ} where the calculation 
of the form factor $\gamma^*\ra \rho_{\perp}+\pi$ is described in detail
\footnote{\hspace{2mm}
The calculation for charmonium is even simpler because the 
three-particle ${\ov c}c g$-wave functions of twist 3 containing the
additional valence gluon have couplings $f_3^i
\sim f_V\Lambda^{2}_{QCD}/{\ov M}$, i.e. are 
suppressed by a factor $\sim (\Lambda_{QCD}/{\ov M})^2$ in comparison to
$f_V{\ov M}\simeq f_P{\ov M}$ in eqs.(4-6), and can be neglected.
}
)\,, and the result for the leading term has the form: 

\bq
|F_{VP}(s)|=\frac{32\pi}{9}\Big |\frac{f_V f_P\ov M}{q_o^4}\Big |\,I_o\,,
\eq
\bq
I_o=\int^1_0 dx_1 \int^1_0 dy_1\ak \Biggl \{\frac{Z_t Z_p 
V_{T}(x) P_{P}(y)}{d(x,y)\, s(x)}-\frac{\ov M_Q^2}{{\ov M}^2}\,
\frac{Z_m^\sigma Z_t V_T(x) P_A(y)}{d(x,y)\,s(x)}+\nonumber
\eq
\bq
+\frac{1}{2}\frac{V_{L}(x)\,P_{A}(y)}{d(x,y)}+\frac{1}{2}
\frac{(1-2y_1)}{s(y)}\frac{V_{\perp}(x)\,P_{A}(y)}{d(x,y)}+
\eq
\bq
+\frac{1}{8} \biggl ( 1-Z_t Z_m^k\frac{4{\ov M}_Q^2}{{\ov M}^2 }\biggr )\, 
\frac{(1+y_1)V_A(x)P_A(y)}
{d^2(x,y)}\Biggr \}.\nonumber
\eq
Here: $s=4E^2,\,q_o^2=(|\vec p|+E)^2\simeq (s-2{\ov M}^2),\, k=(k_1+l_1)$ 
is the gluon momentum in Fig.1,\, $d(x,y)$ and $s(x)$ originate from the 
gluon and quark propagators,\, $Z_{t}$ and $Z_{p}$ are the
renormalization factors of the local tensor and pseudoscalar currents:
\bq
d(x,y)=\frac{k^2}{q_o^2}=\biggl ( x_1+\frac{\delta}{y_1}\biggr )
\biggl (y_1+\frac{\delta}{x_1}\biggr )
\,,\quad \delta=\Biggl (Z_m^k \frac{{\ov M}_Q}{q_o}\Biggr )^2\,, 
\eq
\bq
s(x)= \biggl (x_1+\frac{Z_m^{\sigma}{\ov M}_Q^2}{y_1y_2\,q_o^2} \biggr )
\,,
\quad s(y)= \biggl (y_1+\frac{Z_m^{\sigma}{\ov M}_Q^2}{x_1x_2\,q_o^2}
\biggr )\,,\nonumber
\eq

\bq
\quad Z_p=\Biggl [\az \Biggr ]^{\frac{-3C_F}{b_o}};
 Z_{t}=\Biggl [\az \Biggr ]^{\frac{C_F}{b_o}};
 Z_{m}(\mu^2)=\Biggl [\frac{\alpha_s(\mu^2)}{\alpha_s({\ov M}_Q^2)} 
\Biggr ]^{\frac{3C_F}{b_o}};
\eq
\bq
{\ov M}_Q(\mu^2)=Z_m(\mu^2)\,{\ov M}_Q\,,\quad \quad Z_m^k=Z_m(\mu^2=k^2),\,
\quad Z_m^{\sigma}=Z_m(\mu^2=\sigma^2)\,,
\eq
where ${\ov M}_Q(\mu^2)$ is the running $\ov {MS}$-mass, $C_F=4/3,\,b_o=25/3$.
\vspace{0.5cm}

{\bf 5.} For light quarks the asymptotic forms of wave functions entering 
eq.(4-6) look as follows (see ch.9 and Appendix B in \cite{CZ}).\\
a) for the leading twist 2 wave functions:
\bq
P_A(x)=V_L(x)=V_T(x)=\phi_{asy}(x)=6x_1 x_2\,,
\eq
b) for the non-leading twist 3 wave functions:
\bq
 P_P(x)=1,\quad V_{\perp}(x)=\frac{3}{4}[1+(x_1-x_2)^2]\,,
\eq
\bq
V_A(x)=P_T(x)=6 x_1 x_2\,.\nonumber
\eq

For heavy quarkonium the light-front 
$1S$-Coulomb wave function can be taken as: 
\footnote{\hspace{2mm}
It originates from the Schr{\"o}dinger equal-time wave function
$\Psi_{Sch}(r)\sim \exp(-q_B r)\ra \Psi_{Sch}(\vec k) \sim
(|\vec k|^2+q_B^2)^{-2},$ supplemented with the proposed in \cite{T} and 
commonly used substitution ansatz: 
${\vec k}_{\perp}\to {\vec k}_{\perp},\,\, k_z\to (x_1-x_2) M_0/2,\,\, 
M_0^2=({M_Q^*}^2+{\vec k_{\perp}}^2)/ x_1 x_2\,.$
}
\bq
\Psi (x,\,{\vec k}_{\perp})\sim \Biggl (\frac{{\vec k}_{\perp}^2+
(1-4x_1 x_2){M_Q^*}^2}{4 x_1 x_2}+q^2_B \Biggr ) ^{-2}, \nonumber
\eq
\bq
\phi(x)\sim \int d^2{\vec k}_{\perp}\, \Psi(x,\,{\vec k}_{\perp})\sim
x_1 x_2 \Biggl \{\frac{x_1 x_2}{[1-4x_1x_2(1-{\bar v}^2)]} \Biggr \}\,.
\eq
Here: $q_B$ is the Bohr momentum and ${\bar v}=q_B/M_Q^* \ll 1$ 
is the mean heavy quark velocity. 

For the leading twist charmonium wave functions we will use below a somewhat 
modified simple model form:
\footnote{\hspace{2mm}
It is inspired by the fact that after taking into account small 
relativistic corrections the Coulomb wave function of the $1S$ two-particle 
bound state behaves at small $r$ typically as: $\Psi(r) 
\sim r^{-\Delta}\exp(-q_B r)$. Here $\Delta=c_o {\bar v}^2+O({\bar v}^4)$. 
The constant $c_0$ is not universal and e.g. for the $^1 S_0$\,
-\,states of hydrogen and positronium: $\Delta_{hyd}={\bar v}^2/2+O({\bar v}^
4),\,\,\Delta_{pos}=4{\bar v}^2+O({\bar v}^4)$. Therefore, this can be used 
only at really small ${\bar v}\ll 1$, such that $\Delta\ll 1$, and can't be 
taken literally for charmonium with $v^2\simeq 0.3$. So, we have 
taken the simplest model form of eq.(17) having in mind that it behaves 
qualitatively in a right way and will be really applied at $v^2\simeq 0.3$.
}  
\bq
\phi_{o}(x, v^2)=c_o(v^2)\,x_1 x_2 
\Biggl \{\frac{x_1 x_2}{[1-4x_1x_2(1-v^2)]} \Biggr \}^{1-v^2}\,,
\eq
\bq
\quad \int_0^1 d x_1 \phi_o(x, v^2)=1\,, \nonumber
\eq
where $v$ is now a parameter which has the meaning of the characteristic
quark velocity in the bound state, while $c_{o}(v^2)$ is the 
normalization constant.

The wave function $\phi_o(x, v^2)$ in eq.(17) interpolates in the simplest 
way between two extreme cases: very heavy quark with $v\ra 0$ in 
eq.(16), and very light quark with (formally) $v\ra 1$ in eq.(14).  
In the non-relativistic case of very small $v\ll 1$ the wave function
$\phi_o(x, v^2)$ is strongly peaked around the point $x_1=x_2=1/2$, so that
$\phi_o(x, v^2\ra 0)\ra \delta(x_1-\frac{1}{2})$. And clearly, a decreasing
quark mass leads to larger $v$ and to wider $\phi_o(x, v^2)$.
We take below $v^2= 0.30$, as 
this value is commonly used in calculations for the $1S$ - charmonium.

The wave functions $\phi_o(x,v^2=0.30)\,$ from eq.(17) and 
$\phi_{asy}(x)=\phi_o(x,v^2=1)$\, from eq.(14)
are shown in Fig.2 (the wave function $\phi_o(x, v^2=0.08)\,$ 
corresponding to the scale of $\Upsilon(1S)$ is also shown for comparison 
\footnote{\hspace{2mm}
It is seen from Fig.2 that even $\phi_o(x, v^2=0.08)\,$ is still far from
the $\delta$ - function.}
).

\begin{figure}
\centering
\includegraphics[width=0.99\textwidth]{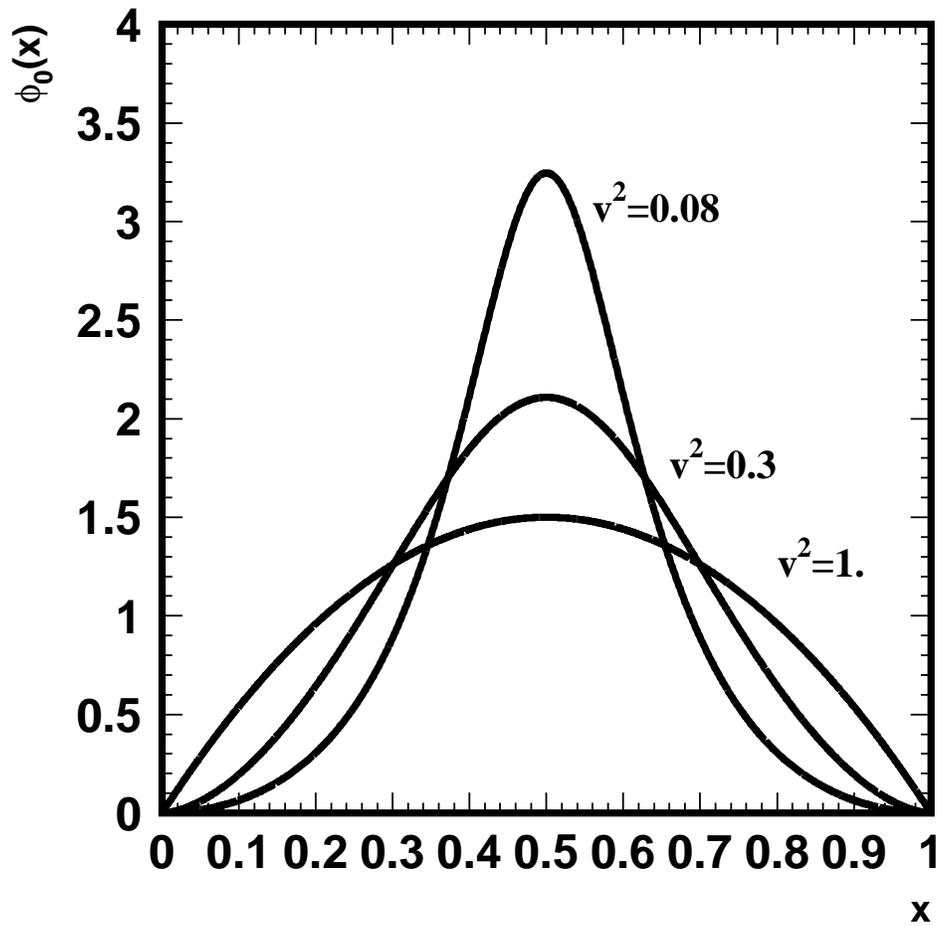}\\

\caption{ The shape of the wave function $\phi_o(x,v^2)$ for: 
$v^2=1$~--~light 
quarks (asymptotic); $v^2=0.30$~--~charmonium; $v^2=0.08$~--~bottomonium.
}
\label{Fig.2}
\end{figure}~

Guided by the above examples, we take for numerical calculations the 
following model wave functions of the $1S$-charmonium:  
\bq
\phi_{i}(x,v^2)=c_{i}(v^2)\,\phi^{asy}_{i}(x)\, \Biggl \{ \frac{x_1 x_2}
{[1-4x_1x_2(1-v^2)]} \Biggr \}^{1-v^2}\,.
\eq

With $v^2=0.30$ this looks as follows (compare to eqs.(14,15),\,
all wave functions are normalized: $\int_0^1 dx\, \phi_{i}(x,v^2)=1$)\,:\\

a) for the leading twist 2 wave functions and $V_A(x)$: 
\bq
V_T(x)=V_{L}(x)= P_A(x)=V_A(x)= 
9.62\,x_1 x_2\Biggl (\frac{x_1 x_2}{1-2.8\, x_1x_2}\,\Biggr )^{0.70}.
\eq 

b) for the non-leading twist 3 wave functions: 
\bq
P_P(x)=1.97\,\Biggl (\frac{x_1 x_2}{1-2.8\,x_1x_2}\,\Biggr )^{0.70},\nonumber
\eq
\bq
V_{\perp}(x)= 1.67\,[1+(x_1-x_2)^2] \Biggl (\frac{x_1 x_2}{1-2.8\,x_1x_2}\,
\Biggr )^{0.70}.
\eq

\vspace{0.5cm}
{\bf 5.} For numerical calculations we use the following parameters. As for 
$\alpha_s(\mu^2)$, we take the simplest form: $(4\pi/b_o)\ln^{-1}(\mu^2/
\Lambda^2)$ with $\Lambda=200\,{\rm MeV}$. It gives, in particular: 
$\alpha_s(\mu^2=M_{\tau}^2)\simeq 0.34$ at the scale of the $\tau$-lepton 
mass, and this looks reasonable. For the ${\ov {MS}}$-mass of the c-quark 
we use the standard value: ${\ov M_c}=1.2\,{\rm GeV}$, and we take 
$f_P\simeq f_{V}\simeq 400\,{\rm MeV}$ on average~
\footnote{\hspace{2mm}
The value of $f_V$ is known from the leptonic decay of $J/\psi$: $\Gamma
(J/\psi\ra e^+e^-)=(16\pi\alpha^2/27)(|f_V|^2/M_{\psi}),\,\,|f_V|\simeq 410
\,{\rm MeV}$. The one-loop corrections $(\sim \alpha_s/\pi)$ to the ratio 
$\Gamma(\eta_c\ra 2\,\gamma)/\Gamma(J/\psi\ra e^+e^-)$ indicate that $f_P$ 
is slightly smaller than $f_V$. 
} .

With all this and the
wave functions given above in eqs.(19,20), one obtains from eqs.(1,9,10): 
\bq
I_o\simeq 5.5\,,\quad
F_{VP}(s=M^2_{\Upsilon(4S)})\simeq 3.5\cdot 10^{-3}\,{\rm GeV}^{-1},\nonumber
\eq
\bq
\sigma \biggl (e^+e^-\ra \gamma^*\ra J/\psi_{\perp}+
\eta_c\biggr )\simeq 33\,{\rm fb}.
\eq

This agrees with the BELLE result (assuming that $Br(\eta_c > 2{\rm charged})
$ is not significantly less than unity). This fact can't be taken too 
literally, of course, due to dependence on the model form of used wave 
functions. Some check of the sensitivity of results to the wave function 
forms can be obtained by variation of the parameter $v^2$ in eq.(18). So, 
we repeated calculations with $v^2=(0.30\pm 0.05)$. Then the cross 
section changes only by $\pm 7\%$. Therefore, the 
results are not very sensitive to reasonable variations of $v^2$.

Now, let us discuss some characteristic features of the above result for the 
integral $I_o$ in eq.(10). The mean value of the renormalization factor $Z_m^
{\sigma}$, see eqs.(12,13), is: $\langle Z_m^{\sigma}\rangle \simeq 0.74$.
This means that in Fig.1 the typical value of the running c-quark 
mass in the quark propagator is: $\langle M_c(\mu^2=\sigma^2)\rangle \simeq 
0.9\,{\rm GeV}$, while $\langle\sigma^2\rangle\simeq 33\,{\rm GeV}^2$ which 
is $\sim 1.5$ times smaller than the typical value $\simeq q_o^2/2
\simeq 46\,{\rm GeV}^2$ for the narrow wave functions.

The value of $\langle k^2\rangle$ can be inferred either from the mean value 
of $\langle Z_m^k \rangle \simeq 0.80$, or from the coupling $\alpha_s(k^2): 
\langle \alpha_s(k^2)\rangle\simeq 0.263$. Both give the same result and show
that the mean virtuality of the gluon in Fig.1 is: $\langle k^2 \rangle\simeq 
12\,{\rm GeV}^2$. This is $\sim 2$ times smaller than a typical rough estimate: 
$\langle k^2\rangle\simeq q_o^2/4\,\simeq 23\,{\rm GeV}^2$ for narrow $\delta$-
like wave functions (and is not so far from the two-quark threshold 
$(2{\ov M}_c)^2\simeq 5.8\,{\rm GeV}^2$).
\footnote{
It is worth noting also that while $f_v^a(\mu^*)$ in eq.(7) is very small
and negative (e.g., using for the c-quark pole mass the value $M^{*}_c=1.65\,
{\rm GeV}$ obtained in \cite{D} from the D-meson semileptonic width, one
obtains: $f_v^a(\mu^*)\simeq -0.08\,),\,\,$ the coupling $f_v^a(k^2)$ from 
eq.(8) becomes positive and much larger, $\langle f_v^a(k^2)\rangle\simeq 
+0.26$, so that the contribution of the term with the wave function 
$V_A(x)$ in eq.(10) is $\simeq 15\%$ of the whole integral $I_o$.
} 

The smaller values of $\langle k^2 \rangle$ and $\langle \sigma^2 \rangle$  
is the main reason why the standard NRQCD-calculations underestimate
the cross section considerably. In other words, the charm quark is not very
heavy and its wave functions are not much like the $\delta$-functions,
although they are of course significantly narrower than similar wave 
functions of really light quarks (see Fig.2).  

It is also of interest to make a comparison to the value of the cross section
obtained from eqs.(1,9,10) in the limit which, in essence, corresponds to the 
approximations of NRQCD, see \cite{BL}). For this, one has to replace in
eq.(10): a) all wave functions, except for $V_A(x)$, by $V_i=P_i=\delta
(x-1/2)$;\, b)~to omit the term with $V_A(x)$;\, b) to replace all $Z_i$ by 1. 
In this case:
\bq
k^2\simeq \big [1+(2\ov M_Q/q_o)^2\, \big ]^2\,\,\frac
{q_o^2}{4}\simeq 26\,{\rm GeV}^2,\quad \alpha_s(k^2)\simeq 0.23\,.
\eq

Substituting all this into eqs.(1,9,10), one obtains:
\bq
I_o\simeq 1.6\,,\quad F_{VP}\simeq  1.1\cdot 10^{-3}\,{\rm GeV}^{-1}\,,\quad
\sigma \simeq 3\,{\rm fb},
\eq
which is essentially smaller than in eq.(21).
\footnote{\hspace{2mm}
In the opposite limit when all $1S$ - charmonium wave functions are taken 
as the asymptotic ones, see eqs.(14,15), the cross section will be $\sigma
\simeq 70\,{\rm fb}$.
}

\vspace{0.5cm}

On the whole, as was argued above, the difficulties with explaining the
BELLE result for $\sigma(e^+e^-\ra J/\psi+\eta_c)$ 
are not really the difficulties of QCD, but are rather  
due to a poor approximation of the real dynamics of c-quarks by
NRQCD. Within the approach described in this paper (which we consider
as more realistic), the BELLE results look rather natural. 

We hope that subsequent experimental and theoretical efforts in this field 
will help to elucidate dynamics and properties of various charmonium
states. In particular, the use of the wave functions $V_i(x)$ given in 
eqs.(19,20) instead of $\delta (x-0.5)$ from NRQCD, will enhance the 
calculated cross sections of inclusive direct production of $J/\psi$. 

\vspace{0.5cm}

We are grateful to A.I. Milstein for explaining us the properties of the
positronium wave functions and a useful discussion.

The work of V.L. Chernyak was supported in part by the RFFI grant 
 03-02-16348-a.

\end{document}